\documentclass[twocolumn]{svjour3}
\usepackage{multirow}
\usepackage{amsmath,amsfonts,amssymb}
\usepackage{setspace}
\usepackage{tocloft}
\usepackage[flushleft]{threeparttable}
\usepackage{natbib}
\usepackage{graphicx}
\usepackage{lipsum}
\usepackage{float}
\usepackage{amsmath}
\usepackage{latexsym}
\usepackage{booktabs,etoolbox}
\usepackage[utf8x]{inputenc}

\newcommand{\as}{\prime\prime}
\renewcommand{\deg}{^{\circ}}
\newcommand{\D}{\Delta}
\renewcommand{\l}{\lambda}
\def\be{\begin{equation}}
\def\ee{\end{equation}}
\def\fr{\frac}
\def\la{\langle}
\def\ra{\rangle}

\newcommand{\apj}{\it Astrophys.~J.}

\newcommand{\apss}{\it Astrophys. Space Sci.}

\begin{document}
\title{Prospect for UV observations from the Moon. III. Assembly and ground calibration of Lunar Ultraviolet Cosmic Imager (LUCI)}
\titlerunning{Assembly and Calibration of LUCI}        

\author{Joice Mathew \and B. G. Nair \and Margarita Safonova \and S. Sriram \and Ajin Prakash\and Mayuresh Sarpotdar \and  S. Ambily  \and Nirmal K. \and A. G. Sreejith \and Jayant Murthy \and P. U. Kamath\and S. Kathiravan \and B. R. Prasad \and Noah Brosch \and Norbert Kappelmann \and Nirmal Suraj Gadde \and Rahul Narayan}

\institute{Joice Mathew \at
Indian Institute of Astrophysics, Koramangala 2nd block, Bangalore, 560034, India and University of Calcutta, Kolkata, 700073, India \\
\email{joicemathewml@gmail.com}
\and B.G. Nair  \and Margarita Safonova\and S. Sriram \and Mayuresh Sarpotdar\and Ajin Prakash \and  S. Ambily  \and K. Nirmal  \and Jayant Murthy\and P.U. Kamath\and S. Kathiravan \and B. R. Prasad \at
Indian Institute of Astrophysics, Koramangala 2nd block, Bangalore, 560034, India                
           \and
           A. G. Sreejith \at Space Research Institute, Austrian Academy of Sciences, Schmiedlstrasse 6, Graz, Austria  \and 
Noah Brosch\at
The Wise Observatory and the Dept. Of Physics and Astronomy, Tel Aviv University, Tel Aviv 69978, Israel\and
Norbert Kappelmann\at
Kepler Center for Astro and Particle Physics, Institute of Astronomy and
Astrophysics, University of T\"{u}bingen, Sand 1, 72076, T\"{u}bingen, Germany
           \and
Nirmal Suraj Gadde \and Rahul Narayan\at
Team Indus, Off Bellary Road, Jakkuru Layout, Bengaluru, 560092, India}

\date{Received: date / Accepted: date}

\maketitle

\begin{abstract}

The Lunar Ultraviolet Cosmic Imager (LUCI) is a near-ultraviolet (NUV) telescope with all-spherical mirrors, designed and built to fly as a scientific payload on a lunar mission with Team Indus -- the original Indian entry to the Google Lunar X-Prize. Observations from the Moon provide a unique opportunity of a stable platform with an unobstructed view of the space at all wavelengths due to the absence of atmosphere and ionosphere. LUCI is an 80 mm aperture telescope, with a field of view of $27.6^{\prime}\times 20.4^{\prime}$ and a spatial resolution of $5^{\prime\prime}$, will scan the sky in the NUV (200--320 nm) domain to look for transient sources. We describe here the assembly, alignment, and calibration of the complete instrument. LUCI is now in storage in a class 1000 clean room and will be delivered to our flight partner in readiness for flight.

\end{abstract}


\section{Introduction}

The Moon has no dearth of scientific instruments, whether on the surface or in orbit, sent there starting from the 1950s: Luna 1, 2, 3 etc., Apollo missions of 1969 to 1972, Chandraayan 1, just to name the few. However, they were all mostly aimed at studying the Moon, or interplanetary space, rather than using it a basis for space observations. Though the Moon was proposed as a prime site for astronomical observations more than 50 years ago \citep{tifft}, the ambitions were always for the large lunar telescopic installations with 4--100 m mirrors, or total mass of 10--30 tons of radio antennas (e.g. \cite{{lunar_radio}, {Angel2008}}), the cost of which much overshot the easier-affordable orbital observatories in the much more accessible Low-Earth Orbit (LEO) environment. 

Moon offers several advantages as a base for space telescopes: stable platform with an unobstructed view of the sky at all wavelengths; essential absence of atmosphere and ionosphere; low gravity; availability of solar power and, if a shielding is provided, constant low temperature allowing deep space observations. The traditionally cited disadvantages are that LEO is easier accessible for repairs and going to LEO is cheaper; and that the orbital space radiation environment is more benign, which also includes micro meteorites. One more challenge for a lunar observatory is considered to be the lunar dust settling on and contaminating telescope optics. However, the only space telescope ever repaired was the LEO Hubble Space Telescope, and the high cost still does not prevent us from launching deep space observatories, such as Spitzer, GAIA, RadioAstron, etc. As for the lunar dust, all the Lunar Ranging Retroreflectors (LRR), including Lunokhod 2’s (placed on the Moon in 1973!), are still functioning on some level (e.g. \cite{currie}). 

Due to the Earth's atmosphere absorbing and scattering UV photons, preventing observations of the active Universe, the Moon's unobstructed sky is especially favourable for the UV observations. Several proposals were put up in the 90s for the UV/optical transit telescope to take advantage of the slow lunar sidereal rate, and for the modest-sized UV/Opt/IR fully-steerable robotic telescopes \citep{{McGraw1994},{Chen1995}}. Several missions have been planned, however, the only realized astronomical UV instruments on the Moon were the far-UV (FUV) camera brought by Apollo 16 team in 1972, and two telescopes from 2014 Chinese Chang’e~3 mission: the Lunar Ultraviolet Telescope (LUT) and the Extreme Ultraviolet camera (EU-VC) \citep{LUT}. The first ever astronomical UV observations from the Moon were performed by the Apollo 16 team 3" Schmidt telescope with FUV camera/spectrograph \citep{{1972Sci...177..788C},{1973ApOpt..12.2501C}}. It could see stars as faint as V magnitude 11, and obtained the first image of the Earth's plasmasphere. Though the Chang'e~3 EUV camera has failed, the LUT -- a robotic 15-cm Ritchey-Chretien NUV telescope -- is still operational, and could do that for 30 years more\footnote{due to the power source for the Chang’e-3 lander. Source: Steve Durst, director of the International Lunar Observatory Association (ILOA) at opening speech of the Global Space Exploration Conference (GLEX) 2017, 6--8 June 2017, Beijing, China.}. With magnitude limit of AB 13 it continues to return significant NUV scientific results in spite of its small size (e.g. \cite{{zhou}, {Meng}}).

Due to the surge in space flights development through both privately funded and newly-developing partnership between private companies and governments, the Moon is revived as a prime place for space astronomy. There are a number of opportunities for flight in the so-called ``new-space'' era, and we became associated with an Indian startup Team Indus\footnote{www.teamindus.in}. As a part of the Google Lunar X-Prize initiative\footnote{https://lunar.xprize.org/}, they planned to launch a mission to the Moon and offered us space on their lunar lander. Though the Google Lunar X-Prize is officially canceled as of now, Team Indus is still going ahead with their mission. Our team has been developing UV payloads \citep{{2018ExA....45..201M},{2018SPIE10699E..3EA},{2016SPIE.9908E..4EA},{2015SPIE.9654E..0DS}} ready to fly on a range of available platforms, and LUCI was our choice for the lunar mission. 

\begin{figure*}[ht]
\begin{center}
\includegraphics[scale=0.85]{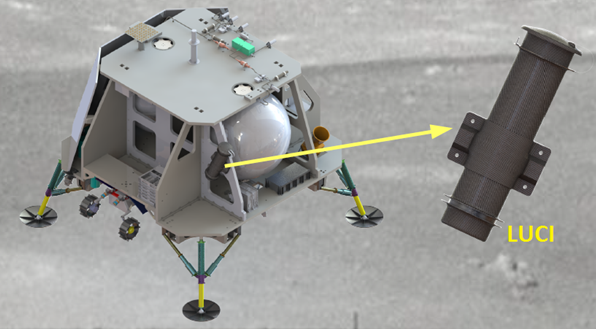}
\end{center}
\caption{LUCI on the Team Indus lunar lander.}
\label{fig:LUCI on the TeamIndus lunar lander}
\end{figure*}

The original design of the instrument was the FUV re-orientable telescope (\cite{Safonova}, hereafter paper~I); however challenges due to constraints on weight, volume, and power placed by the spacecraft team forced us to significantly reduce the size and weight of the payload as well as change the wavelength band from FUV to NUV. The mechanical constraints along with budgetary restrictions, availability of space-qualified components (detector \& optics) and time-bound development resulted in the new instrument, where it was the design concept that defined the science goals (\cite{Mathew}, hereafter paper~II). LUCI is proposed to be mounted on the Team Indus lander at fixed angle (Fig.~\ref{fig:LUCI on the TeamIndus lunar lander}) and scan the sky with the lunar rotation with a field of view (FOV) of $0.34^{\circ}\times 0.46^{\circ}$, detecting bright variable UV sources such as variable stars, novae, M-dwarf flares etc. The details of scientific objectives and the observational strategy of LUCI are presented in paper~II.

In this paper, we describe the steps involved in the opto-mechanical assembly of the telescope, ground tests and calibrations results of LUCI, and the achieved instrument performance. The instrument assembly and calibration were performed at the M.~G.~K.~Menon Laboratory for Space Sciences at the CREST campus of the Indian Institute of Astrophysics (IIA), Bengaluru \citep{kumar}, and LUCI is currently stored in a clean room in ultra-pure nitrogen atmosphere.

\section{Instrument Overview}

\begin{figure*}[ht]
\begin{center}
\includegraphics[scale=0.18]{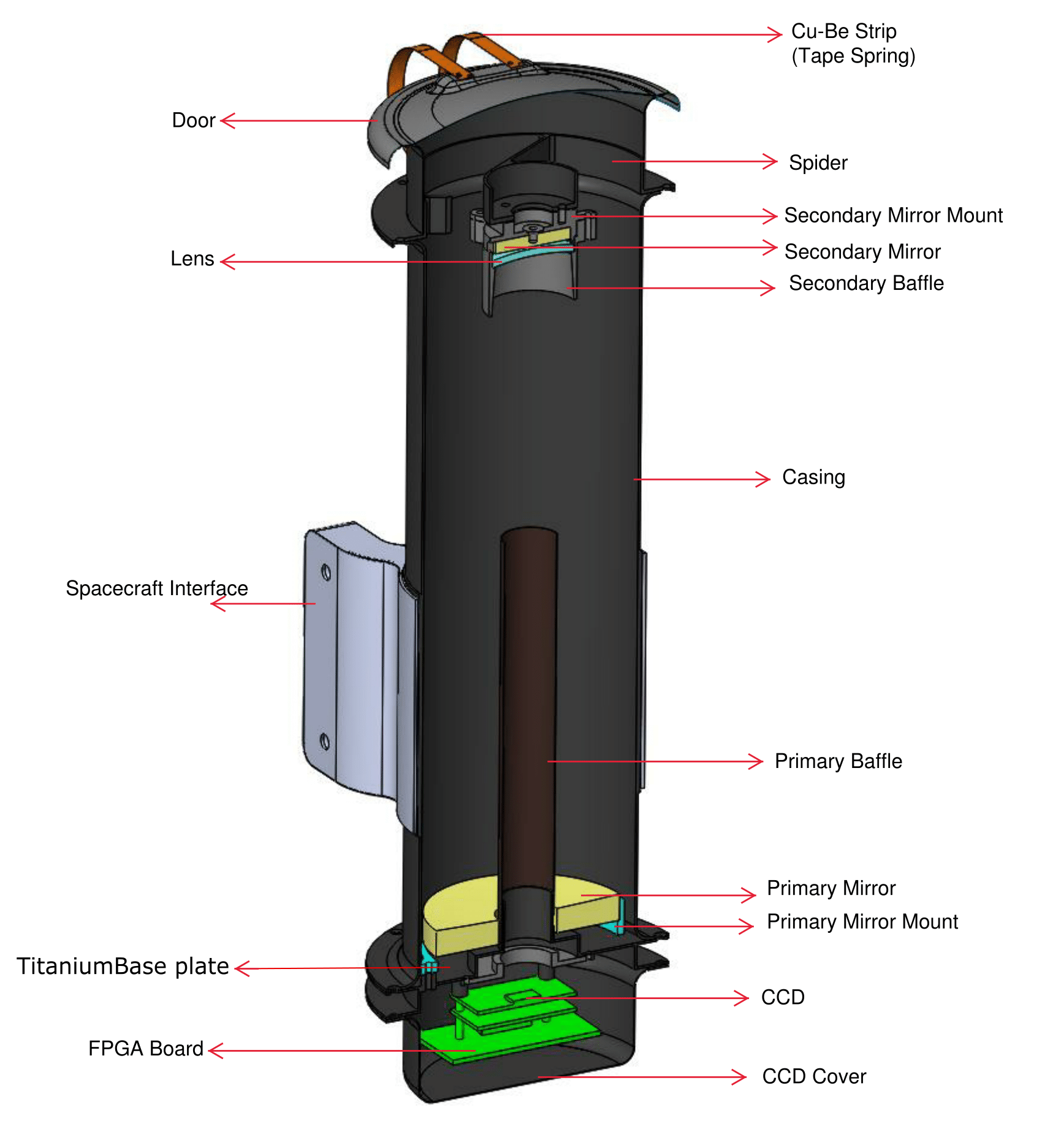}
\end{center}
\caption{LUCI cross-section}
\label{fig:LUCI cross-section}
\end{figure*}

Detailed instrument design of LUCI is described in paper~II, and the basic technical specifications are given in Table~\ref{tab:specs}. The cutaway of the LUCI instrument is shown in Fig.~\ref{fig:LUCI cross-section}. The structure is made of carbon fiber reinforced polymer (CFRP) cast as a single tube; the telescope will be attached to the main body of the lander through a CFRP interface plate at an angle of $25^{\circ}$ from zenith to avoid Sun and the horizon glow (the angle may change if the lander is redesigned). A single-use door protects the optics from contamination on the ground and from the initial puff of dust, expected upon landing on the lunar surface. The door will be opened to begin the observations two days after the  landing, which should provide sufficient time for the dust (due to landing) to settled down. The opening mechanism employs the nichrome wire, nylon rope and beryllium-copper spring; current passed through the nichrome wire melts the rope and releases the spring, opening the door. 

\begin{figure}[ht]
\includegraphics[scale=0.43]{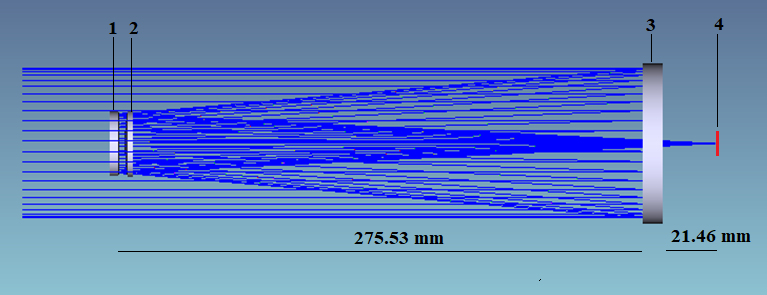}
\caption{Optical layout: 1. secondary mirror; 2. corrector lens; 3. primary mirror; 4. focal plane.}
\label{fig:Optical Layout}
\end{figure}

The optical layout of LUCI is shown in Fig.~\ref{fig:Optical Layout}. The micro roughness requirement for the LUCI optics is 25 \r{A} rms. The corrector lens is placed in front of the secondary mirror to reduce the aberrations from other spherical elements. The incident light is focused on a UV-enhanced CCD at the focal plane at a distance of $310$ mm from the lens. 

\begin{table}[h]
\caption{Technical details}
\begin{center}
\begin{tabular}{ll}
\hline 
\rule[-1ex]{0pt}{3.5ex} Instrument & UV Imager(LUCI) \\\hline
\rule[-1ex]{0pt}{3.5ex} Telescope type & Cassegrain  \\
\rule[-1ex]{0pt}{3.5ex} Primary mirror (PM) &  80-mm dia      \\
\rule[-1ex]{0pt}{3.5ex} Secondary mirror (SM) &   32.4-mm dia       \\
\rule[-1ex]{0pt}{3.5ex} Field of view   &  $27.6^{\prime}\times 20.4^{\prime}$  \\
\rule[-1ex]{0pt}{3.5ex} Focal length & 800.69 mm  \\
\rule[-1ex]{0pt}{3.5ex} Spatial resolution & $5^{\prime\prime}$  \\
\rule[-1ex]{0pt}{3.5ex} Detector & UV-enhanced CCD \\
\rule[-1ex]{0pt}{3.5ex} Sensor format (H$\times$V)& $1360 \times 1024$ pixels\\
\rule[-1ex]{0pt}{3.5ex} Pixel size & $4.65\times 4.65\,\mu$m \\
\rule[-1ex]{0pt}{3.5ex} Wavelength band & 200 -- 320 nm\\
\rule[-1ex]{0pt}{3.5ex} Time resolution & 0.08 sec (12 fps)\\
\rule[-1ex]{0pt}{3.5ex} Exposure time (minimal) & $2309$ sec \\
\rule[-1ex]{0pt}{3.5ex} Power & $ < 5$  W \\
\rule[-1ex]{0pt}{3.5ex} Dimension (L $\times$ D)& $450 \times 150$ mm \\
\rule[-1ex]{0pt}{3.5ex} Weight & $ 1.2 $ kg \\
\hline
\end{tabular}
\label{tab:specs}
\end{center}
\end{table}

The detector has a non-negligible response in the range of 200--900 nm, therefore we have placed a solar blind filter before the CCD to restrict the bandpass to 200--320 nm (see Fig.~11 in paper~II for both curves). The detector electronics includes a generic in-house developed field-programmable gate array (FPGA) board used as the image processor board to generate the clocks and read the data, and a real-time processor system for image processing tasks of different levels \citep{fpga-board}. The electronic components are shielded with Aluminum to mitigate the cosmic rays hits. 

Primary and secondary baffles are implemented in LUCI to reduce the effect of the stray light at the focal plane from both the Sun and the Earth (scattered light is suppressed to the order of $10^{-12}$ for a light source at $45^{\circ}$ from LUCI optical axis; Fig.~6 in paper~II). The inside walls of the telescope tube, mirror mounts and baffles are black painted with Aeroglaze\textsuperscript{\textregistered} Z306 to suppress the scattering. 

\section{Opto-mechanical assembly and alignment}

\subsection{Optics and mount interface}

\begin{figure}[ht]
\begin{center}
\includegraphics[scale=0.75]{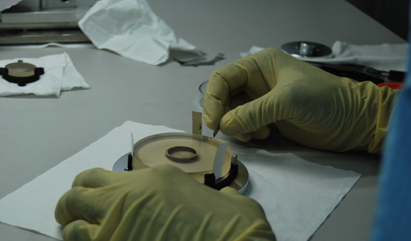}
\end{center}
\caption{Gluing of optics to the Invar mount}
\label{fig:optics-mount assembly}
\end{figure}

We began the assembly process by attaching the primary mirror to its mount: mirror was placed on the three arms of the mount and its orientation was adjusted using $100\,\mu$m shims at each interface. The mirror was finally secured on the mount using a space-grade adhesive (3M\textsuperscript{\textregistered} 2216 B). Each of the three blade projections of the Invar mount has two pinholes to expel the excess glue during the bonding and curing process. Gluing of the optics was performed in a class 10000 clean room environment and, after gluing, all components were left undisturbed for seven days for curing. A portion of the same glue was kept aside in an aluminium foil in identical conditions to verify the strength of the bonding. The same procedure was repeated for the secondary mirror. The process of gluing the primary mirror onto the Invar mount is shown in Fig.~\ref{fig:optics-mount assembly}.

The Primary mirror mount along with the mirror was then connected to the titanium base plate -- an interface between the primary mirror cell and the detector mounting -- by titanium M3 bolts at three points $120\deg$ apart. The bolts were torque-tightened using a torque wrench, restricting the maximum torque to 80\% of the maximum allowed value. 

\subsection{Tolerance Requirements}
\begin{figure*}[h!]
\begin{center}
\includegraphics[scale=0.55]{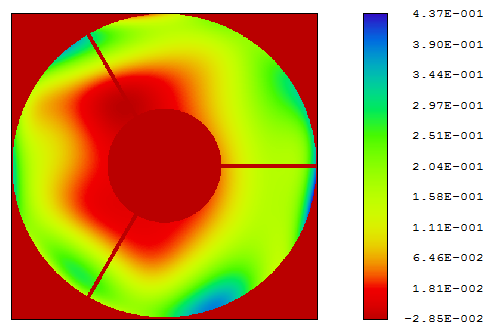}
\end{center}
\caption{Wave Front map based on the tolerance analysis}
\label{fig:Wave Front Error map}
\end{figure*}

The scientific requirement of LUCI is to have a minimum spatial resolution of $5^{\as}$. We have performed the sensitivity tolerance analysis to derive the alignment and manufacturing tolerance parameters for different components of LUCI to achieve the scientific requirement.  The requirements for alignment were obtained from tolerance analysis performed on Zemax\textsuperscript{\textregistered}. The input values used were the tolerance values provided by the manufacturer, and those imposed by the actual alignment procedure of each element (Table~\ref{table:tolerance values}). Based on these values, we obtained the root-mean-square (RMS) wavefront error of $0.066\,\mu$m as the optical performance limit. The simulated wavefront map based on the tolerance analysis is shown in Fig.~\ref{fig:Wave Front Error map}. The final system-level alignment requirements for LUCI are as follows:
\begin{itemize}
\item{The final RMS WFE of the system should be below $0.066\,\mu$m.}
\item{The detector center has to be aligned with the optical axis within $10^{\as}$.}
\item{Detector position w.r.t. telescope focus shall be within $100\,\mu$m.}
\item{Optical axis and mechanical axis have to match within $30^{\as}$.}
\item{The FWHM of the point spread function (PSF) shall fall   within $5\times5$ pixels (corresponds to $\sim 6^{\as}$).}
\end{itemize}

\begin{table*}[ht]
\begin{center}
\caption{Tolerance allocation on manufacturing and alignment}
\begin{tabular}{lllc}
\hline
Tolerance term & sub tolerance term & objects & tolerances\\
\hline
\rule[-1ex]{0pt}{3.5ex} Manufacture & Radius of curvature (\%) & Primary Mirror & 1 \\
   &  & Secondary Mirror & 1 \\
  &  & Lens & 0.1 \\
  & Thickness ($\mu$m) & Primary Mirror & $\pm 100$\\
  &  & Secondary Mirror &$\pm 100$ \\
  &  & Lens & $\pm 50$ \\
  & Decenter in X $\&$ Y ($\mu$m) & Primary Mirror & $\pm 50$\\
  &  & Secondary Mirror &$\pm 50$ \\
  &  & Lens & $\pm 50$ \\
  & Tilt in X $\&$ Y ($^{\prime\prime}$)  & Primary Mirror & 60\\
  &  & Secondary Mirror &60 \\
  &  & Lens & $60$ \\
  & Surface accuracy ($\mu$m) & Primary Mirror & .01\\
  &  & Secondary Mirror &.01 \\
  &  & Lens & .01 \\
\rule[-1ex]{0pt}{3.5ex} Alignment & Decenter in X $\&$ Y ($\mu$m) & Primary Mirror & $\pm 50$\\
            &  & Secondary Mirror &$\pm 50$ \\
            &  & Lens & $\pm 50$ \\
             & Tilt in X $\&$ Y ($^{\prime\prime}$)  & Primary Mirror & 60\\
             &  & Secondary Mirror &60 \\
             &  & Lens & $60$ \\
\hline
\end{tabular}
\label{table:tolerance values}
\end{center}
\end{table*}

\subsection{Optical alignment}

\begin{figure*}[h!]
\begin{center}
\includegraphics[scale=0.4]{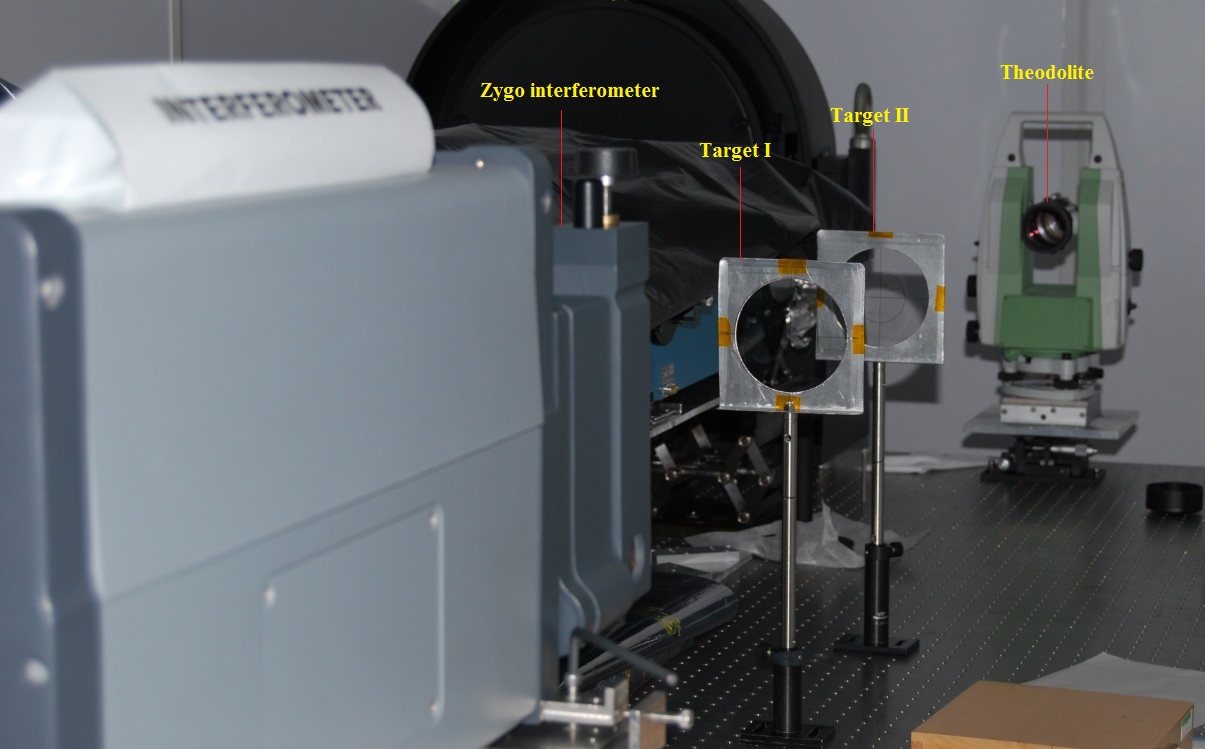}
\end{center}
\caption{Setup for optical axis establishment.}
\label{fig:zygo_theodolite}
\end{figure*}

\paragraph{\bf Coarse Alignment}
The coarse alignment of LUCI has been carried out using a theodolite with cross-hair targets. The theodolite has a precision level of $0.01^{\as}$. We first established an optical axis by mounting all the elements (theodolite, Zygo interferometer, and LUCI) on a zero-vibration precision-controlled optical table (Fig.~\ref{fig:zygo_theodolite}) and setting the theodolite at elevation angle of $90\deg$. We then aligned the theodolite to a $100\,\mu$m pinhole mounted on the aperture of the Zygo interferometer. We defined the optical axis with respect to two transparent targets (Target~I and Target~II) inscribed with cross-hairs placed between interferometer and theodolite, and  adjusted the tip and tilt of the interferometer to ensure that the light beam passes through the centres of the cross-hairs of both targets. Once the optical axis was established, the elevation and azimuthal angles of the theodolite were kept at $90\deg$ and $0\deg$, respectively.

The telescope tube with only primary mirror was mounted onto a custom-made $XYZ$ stage through the spacecraft interface flange. The assembly was fixed on the optical table, replacing Target~II. We used the theodolite to match the optical axis of the primary mirror with the mechanical axis of the telescope tube. We measured the relative tilt and decentre between the primary mirror and front end the telescope tube, and estimated the required thickness of the shims. The final accuracy achieved in the alignment of the optical and mechanical axes was $4^{\as}$. 

In the second step, the optical axis of the primary mirror was matched with the theodolite optical axis by placing the Target~I  at a distance from the primary mirror equal to its radius of curvature.

Next step was mounting the secondary mirror along with the corrector lens on the spider unit. The spider unit was then attached to the telescope tube by placing nuts and bolts at six interface points. 
We have made an aluminium mask/plate with a $50\,\mu$m hole at the center of the plate with positional accuracy of $5\,\mu$m w.r.t. the edges of the plate. This plate was mounted at the place where the CCD printed circuit board (PCB) was intended to be mounted.

\begin{figure*}[h!]
\begin{center}
\includegraphics[scale=0.15]{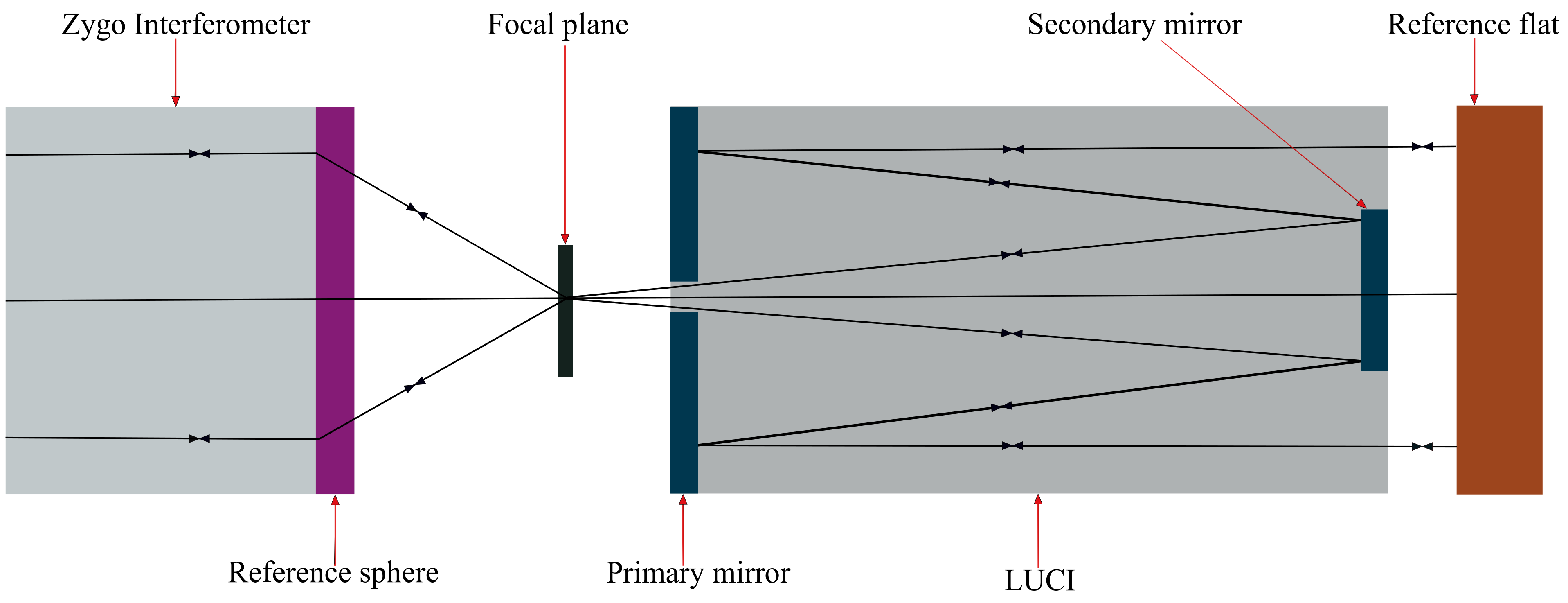}
\end{center}
\caption{Schematics of the fine alignment setup.}
\label{fig:luci final alignment set up block diagram}
\end{figure*}

\begin{figure*}[h!]
\begin{center}
\includegraphics[scale=0.45]{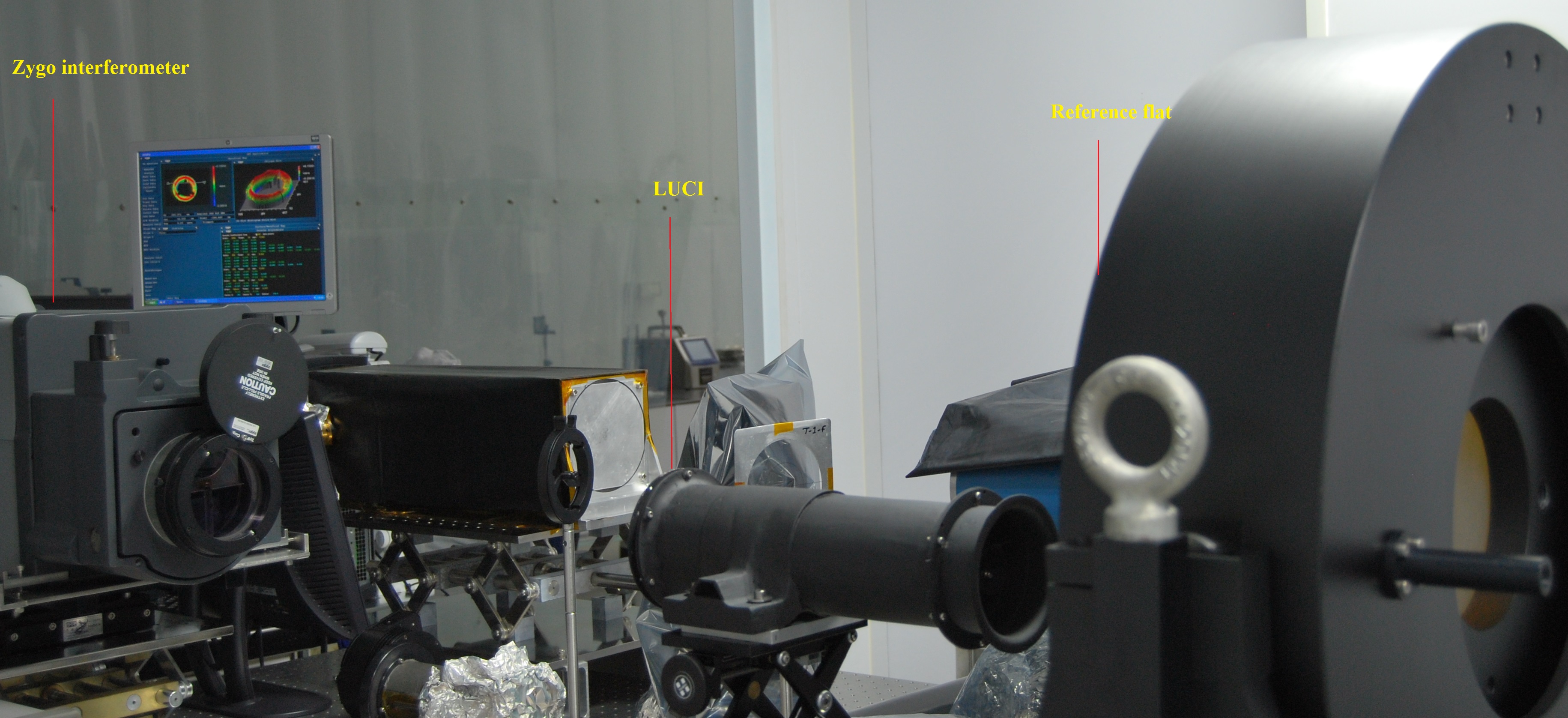}
\end{center}
\caption{LUCI interferometric alignment set-up using Zygo interferometer and reference flat.}
\label{fig:luci final alignment set up}
\end{figure*}

\paragraph{\bf Fine Alignment} The fine alignment of LUCI was performed using the Zygo interferometer. We attached the F7 reference sphere in front of the Zygo interferometer, which focussed the collimated beam onto the Target~I (place in LUCI's focal plane). A very high precision surface flat with surface accuracy of $\Lambda/6$ peak-to-valley, is used in the fine optical alignment set up of LUCI. The beam from the interferometer is reflected through the telescope and the reference flat \footnote{from Optical Surfaces Ltd. (England)http://www.optisurf.com.} placed in front of the entrance aperture of the telescope (see schematics in Fig.~\ref{fig:luci final alignment set up block diagram}). The reference flat reflects back the collimated beam back to its focus and then to Zygo. The returned beam and incident beam interfere and generate interference fringes. The experimental setup for the final alignment is shown in Fig.~\ref{fig:luci final alignment set up}. Resulted fringes obtained from the interferogram were analyzed using the Metro Pro software package and fringe coefficients were derived. 

\subsection{Optical Performance}

\begin{figure*}[ht]
\begin{center}
\includegraphics[scale=0.75]{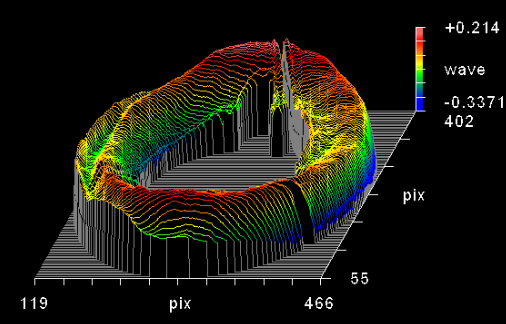}
\end{center}
\caption{Measured WFE map of the final aligned telescope.}
\label{fig:WFE_measured_final_align}
\end{figure*}

From derived fringe coefficients we found that the aberrations were dominated by coma and astigmatism. We corrected the tilt and decenter of the SM to achieve the minimum aberration, and locked its position by torque-tightening the fasteners at the six mounting position. The total optical performance of the telescope after alignment was estimated by analysing the Wave Front Error (WFE) (Fig.~\ref{fig:WFE_measured_final_align}). WFE was determined as 1/5-wave, equivalent to 53 nm.

\subsection{Final Assembly}

\begin{figure*}[ht]
\includegraphics[scale=0.89]{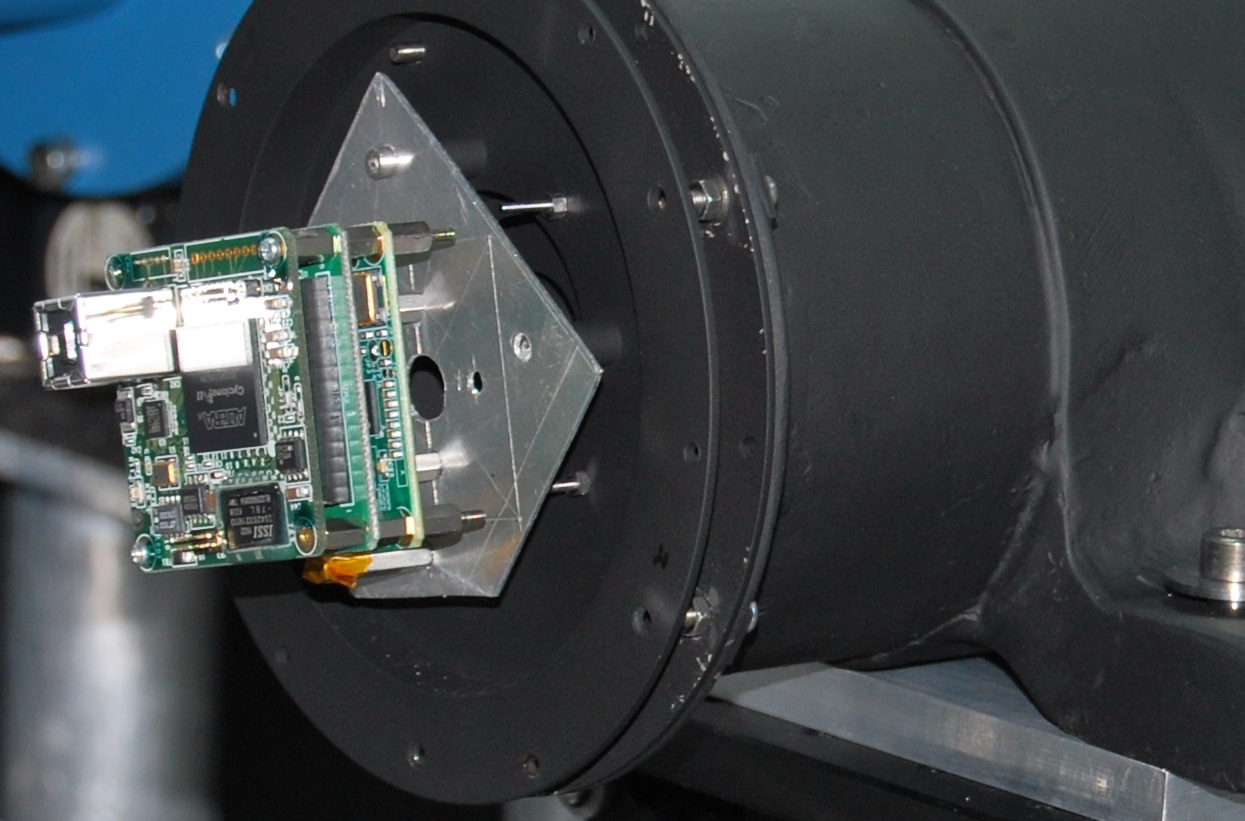}
\hskip 0.3in
\includegraphics[scale=0.35]{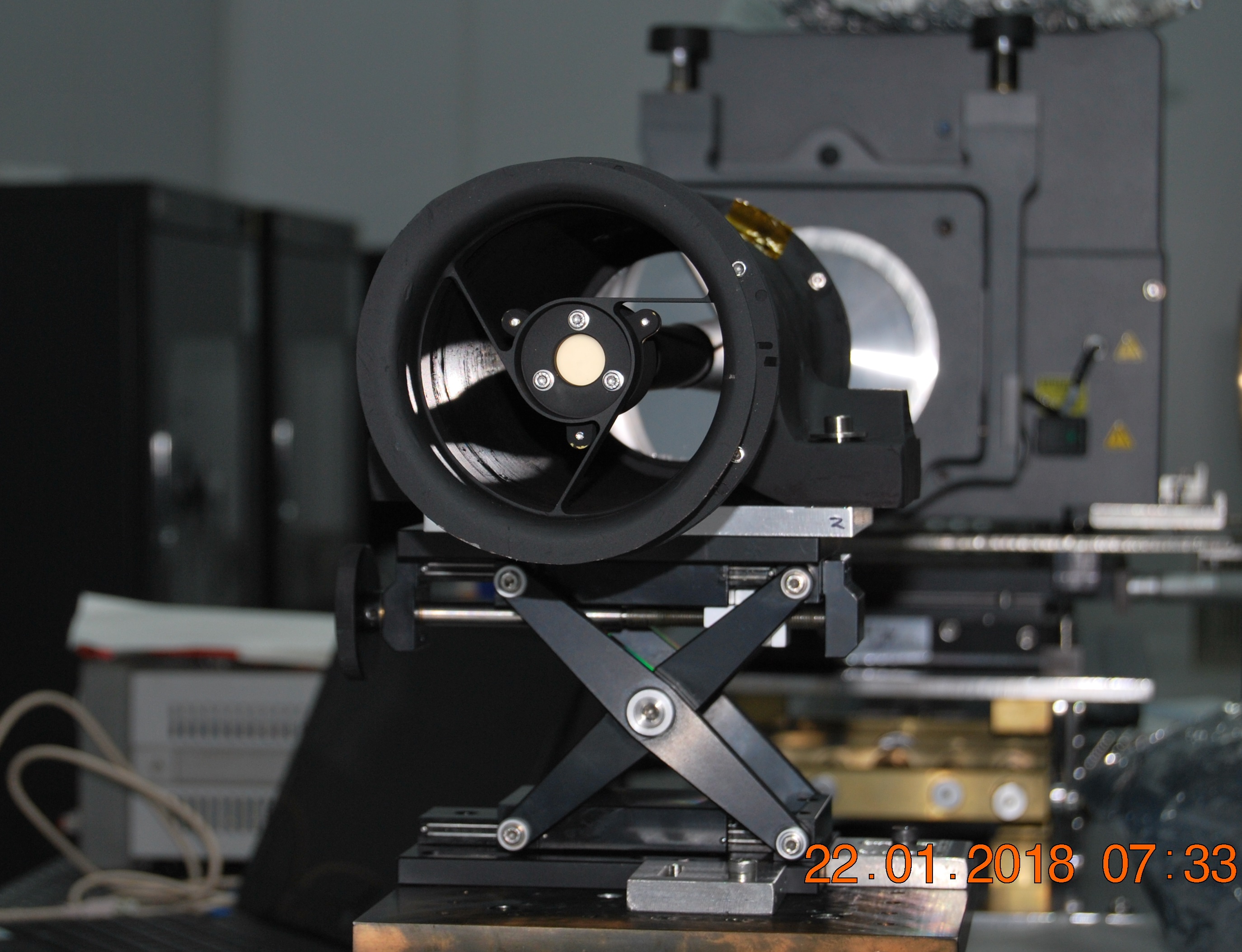}
\caption{{\it Left}: The detector and readout electronics PCB assembly at the back side of the primary mirror. {\it Right}: LUCI after the final opto-mechnical assembly in class 1000 clean room at the M.G.K. Menon Laboratory for Space Sciences.}
\label{fig:assembly}
\end{figure*}

Finally, we fixed the detector PCB at the four interface points ensuring by the use of spacers that the CCD is exactly at the focal plane of the system. This was further verified by imaging a pinhole located at the focal plane of the UV collimator. The detector and the readout electronics assembly at back side of the PM is shown in Fig.~\ref{fig:assembly} ({\it Left}). The final opto-mechanical assembly of LUCI in the class 1000 clean room at the M.G.K. Menon Laboratory for Space Sciences is shown in Fig.~\ref{fig:assembly} ({\it Right}).

\section{Contamination control}

As with all UV payloads, contamination is a critical area of concern in the performance of the instrument. We have therefore performed the assembly and integration of LUCI in a Class 1000 environment with requirements on any material of a TML (Total Mass Loss) value less than 1\%, and a CVCM (Collected Volatile Condensible Mass) value less than 0.1\%. In addition, we adopted the following cleaning process before taking any mechanical piece into the clean room:
\begin{itemize}
\item{Vacuum cleaning.}
\item{Solvent wiping with acetone.}
\item{Ultrasonic cleaning in acetone bath.}
\item{Isopropyl alcohol (IPA) wiping with clean tissues.}
\end{itemize}

LUCI structural parts have been painted black using Aeroglaze\textsuperscript{\textregistered}. After this, they were ultrasonically cleaned in an acetone bath, wiped with isopropanol (IPA), and baked at $100\deg$C in a high vacuum (10$^{-4}$ mbar) chamber for 72 hours. This procedure tagged them as precision-cleaned, and care was taken to maintain their cleanliness. The contamination was monitored with MgF$_2$ witness sample windows kept as close to the critical hardware as possible. The window transmission was checked periodically observing a total variation of less than 5\% over a period of 6 months. Particulate contamination was monitored by the UV (370--390 nm) inspection and by using the particle counters. The particle count in the clean room has been verified on a daily basis.

\section{Tests and calibration}

\subsection{UV Collimator}

\begin{figure*}[h!]
\begin{center}
\includegraphics[scale=0.6]{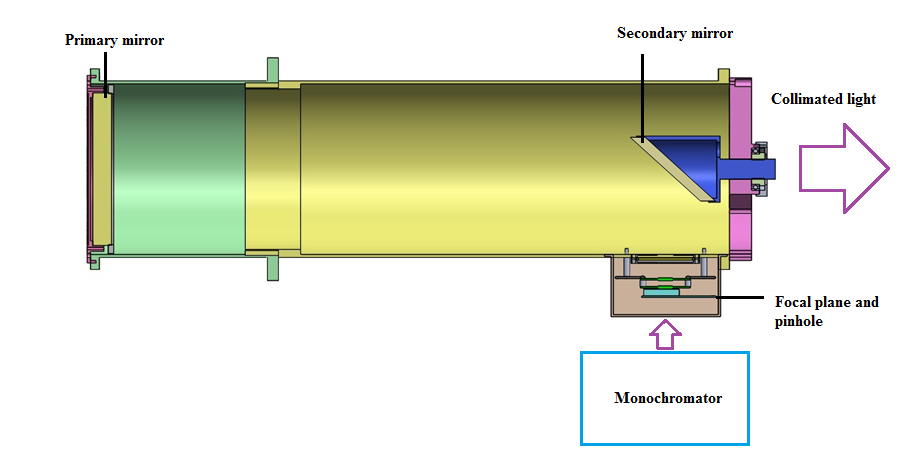}
\end{center}
\caption{UV Collimator schematic, including made in-house Newtonian telescope.}
\label{fig:UV Monochromator}
\end{figure*}

\begin{figure}[h!]
\begin{center}
\includegraphics[scale=0.4]{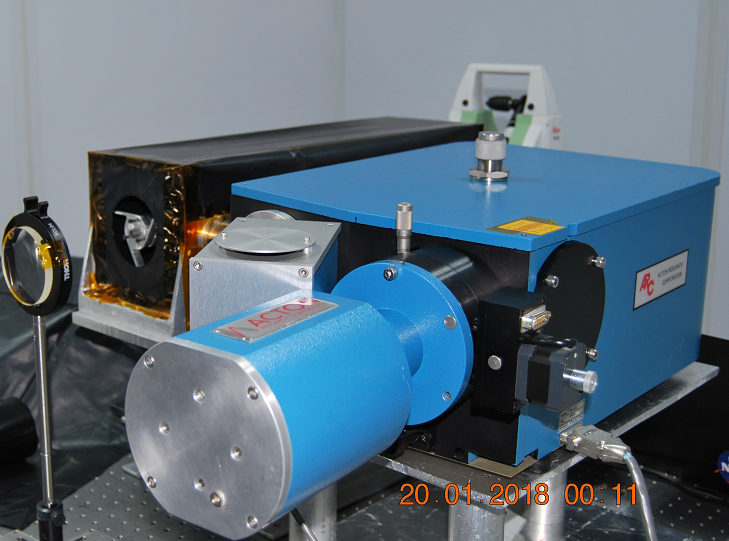}
\end{center}
\caption{UV collimator setup. The whole setup was covered with black paper during the calibration.}
\label{fig:UV collimator}
\end{figure}

We built an F4, 600-mm focal length Newtonian telescope (see Fig.~\ref{fig:UV Monochromator}) to provide the collimated light to the entrance aperture of LUCI. Assembly, alignment and calibration of the collimator were carried out in the clean room, using the techniques described in previous sections. The collimator has a 150-mm diameter parabolic primary mirror and an elliptical flat secondary mirror with a 35-mm minor axis diameter, chosen to have a projected size smaller than the size of LUCI SM. The collimator's plate scale is $343^{\as}$/mm, and the resultant RMS WFE of the collimator telescope was 15 nm.
Light through a $5\,\mu$m-dia pinhole at the focal plane of the collimator passes through an 80-mm aperture mask at the output of the collimator, and the resultant collimated beam is imaged onto the LUCI entrance aperture. The monochromator (SpectraPro\textsuperscript{\textregistered}-300 from Acton Research Corporation) with a deuterium light source is used to illuminate the pinhole in the 200--320 nm wavelength range. For calibration, the whole setup was covered in black paper to avoid a light leak. The UV collimator setup in class 1000 clean room is shown in Fig.~\ref{fig:UV collimator}.

\subsection{Detector characterization}

We are using a broadband Sony ICX407BLA CCD, specially enhanced for the UV response. This is a front-illuminated CCD with quartz window coated with AR Lumogen, a composite phosphor coating,  developed by Photometrics and Princeton Instruments, that improves the sensitivity of CCDs in blue-visible and UV. Additional bonus is that such AR coating has been reported to actually improve the UV efficiency of CCDs in operation under high vacuum (e.g. \citet{eso}). The CCD is a diagonal 8-mm (type 1/2-inch) interline solid-state image sensor with $1360\times 1024$ pixel format and $4.65\,\mu$m pixel size. We have developed a generic field-programmable gate array (FPGA) board \citep{fpga-board} to generate clocks, read the CCD digital output data, and perform on-board data processing. Detector ambient temperature on-board the lander will be maintained between $19^{\circ}$C and $23^{\circ}$C by using an active thermal control system, which consists of a closed-loop control system with heaters, optical solar reflectors (OSR) and a thermistor. LUCI will be covered with the multi-layer insulation (MLI) to achieve the thermal insulation.

We have estimated the dark current of the CCD from a series of dark frames taken at different exposure times. The time-dependent dark current is estimated to be 1.2 e/pixel/sec at $23\deg$C. Readout noise was measured using a large sample of bias images associated with the dark exposures. The average readout noise was measured to be 9 e RMS.

\subsection{Quantum efficiency}
\label{sec:QE}

To measure the quantum efficiency (QE) of the detector, we placed LUCI in front of the UV collimator. The monochromator was turned on for more than two hours before the measurement to stabilise the source output. We acquired the images of the pinhole at 10-nm intervals in 200--320 nm wavelength range. 

\begin{figure}[h]
\begin{center}
\includegraphics[scale=0.75]{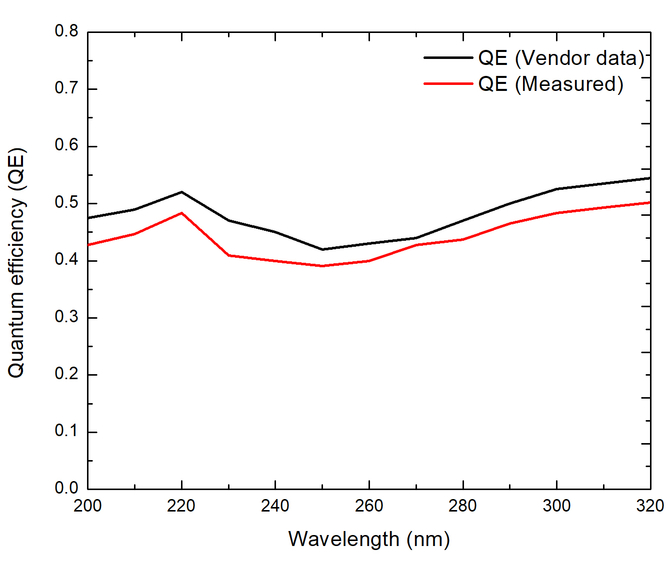}
\end{center}
\caption{Quantum efficiency of the detector.}
\label{fig:Quantum_Efficiency}
\end{figure}
After this, we replaced the CCD PCB with a NIST (National Institute of Science and Technology) calibrated photodiode, and measured the current output corresponding to 200--320 nm flux level. We found the total counts from the pinhole using aperture photometry, and derived the total electrons generated for each 10 nm wavelength flux level. We derived the input flux to the CCD by converting the current measured by the NIST photodiode to photons. The QE, defined as the ratio of the detected electrons to the input flux, is plotted in Fig.~\ref{fig:Quantum_Efficiency}.

\subsection{Field of View (FOV)}

\begin{figure}[ht]
\includegraphics[scale=0.37]{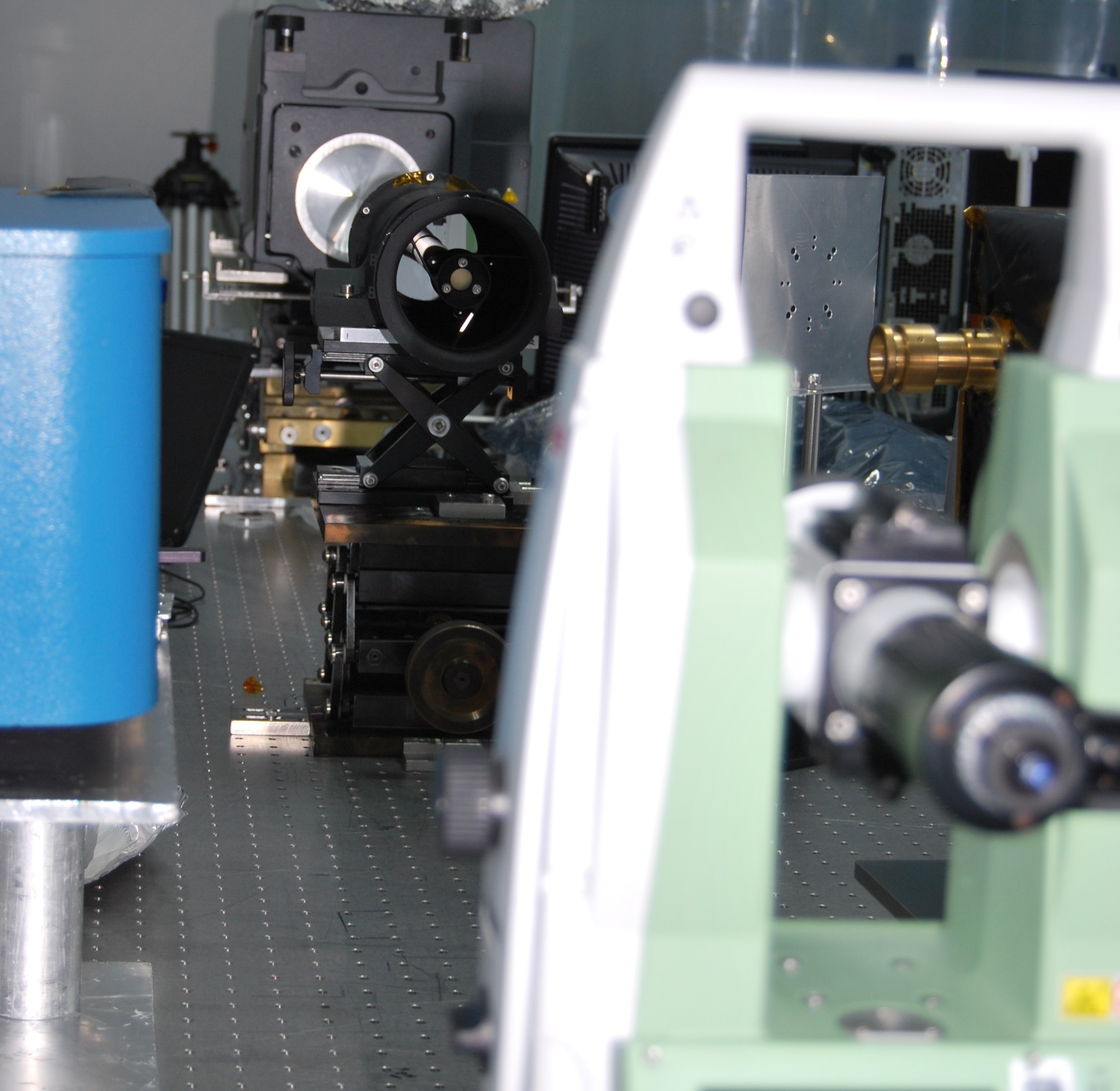}
\caption{FOV measurement setup.}
\label{fig:Field of view measurement set up}
\end{figure}

To measure the FOV, we have made a transparent cross-hair target with a rectangular-shaped mask at the centre, with the size of the mask equal to the CCD's active area ($6.4 \times 4.8$ mm). We placed the mask at the focal plane of LUCI and illuminated it through the telescope. We observed the mask using the theodolite in auto-collimation mode, and moved the theodolite both in azimuth and elevation to find the edges of the mask. This gave us the instrument FOV as $27.55^{\prime}\times 20.37^{\prime}$. The FOV measurement setup is shown in Fig.~\ref{fig:Field of view measurement set up}.

\subsection{Plate scale}

By shifting LUCI on the optical table perpendicular to the theodolite optical axis by a fixed distance, we measured the position of a 5 $\mu$m-size pinhole image on LUCI CCD. The accuracy of the position on the optical table is 10 $\mu$m, and the accuracy of the theodolite pointing is $0.01^{\as}$. By measuring the relative position of LUCI using theodolite and the shift in the pinhole image, we derived the plate scale as $1.21^{\as}$/pixel.

\subsection{Point Spread Function (PSF)}

\begin{figure}[h!]
\includegraphics[scale=0.48]{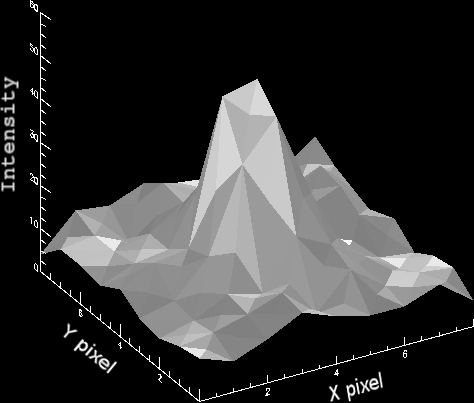}
\caption{3-D plot of LUCI's image (1.4 magnification) of 5-$\mu$m pinhole in the UV collimator focal plane.}
\label{fig:luci psf}
\end{figure}

The measurement of the PSF was conducted by exposing LUCI to the image of a $5\,\mu$m pinhole observed through the collimator. The measurements were ultimately limited by the figure quality of the collimator primary mirror, which has some low-level defects. To measure the off-axis PSF, we tilted the telescope with respect to the central axis to a known distance, and measured the PSF at the edges of the FOV. The on-axis measured FWHM is 3.85 pixels, and the off-axis (edge of the field $\sim 13.8^{\prime} $) FWHM is 4.25 pixels. The 3D plot of LUCI's image (with 1.4 magnification) of the $5\,\mu$m pinhole in the UV collimator focal plane is shown in Fig.~\ref{fig:luci psf}.

\subsection{Filter transmission}

\begin{figure}[h!]
\hspace{-0.25in}
\includegraphics[scale=0.36]{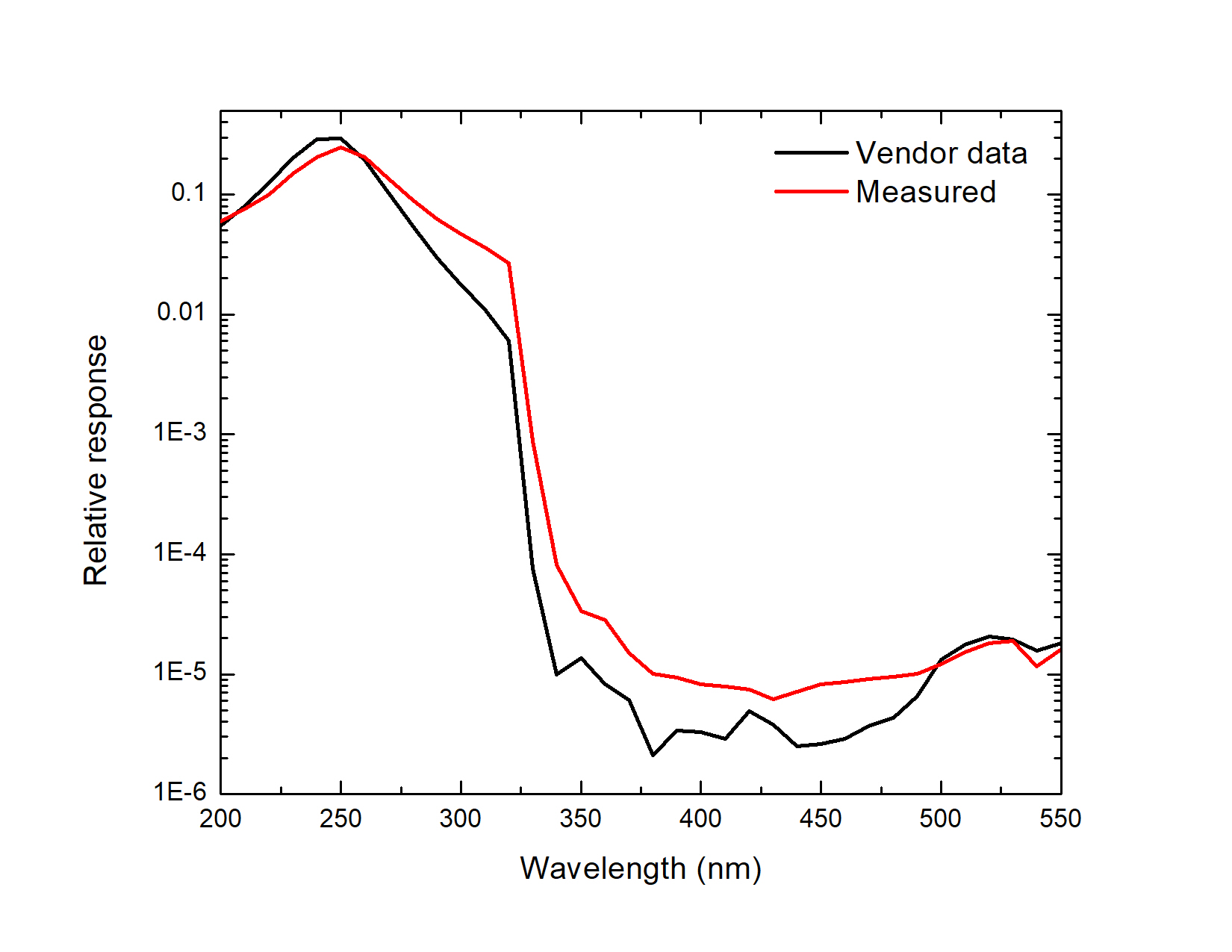}
\caption{Filter transmission}
\label{fig:Filter transmission}
\end{figure}

\begin{figure}[h!]
\hspace{-0.05in}
\includegraphics[scale=0.37]{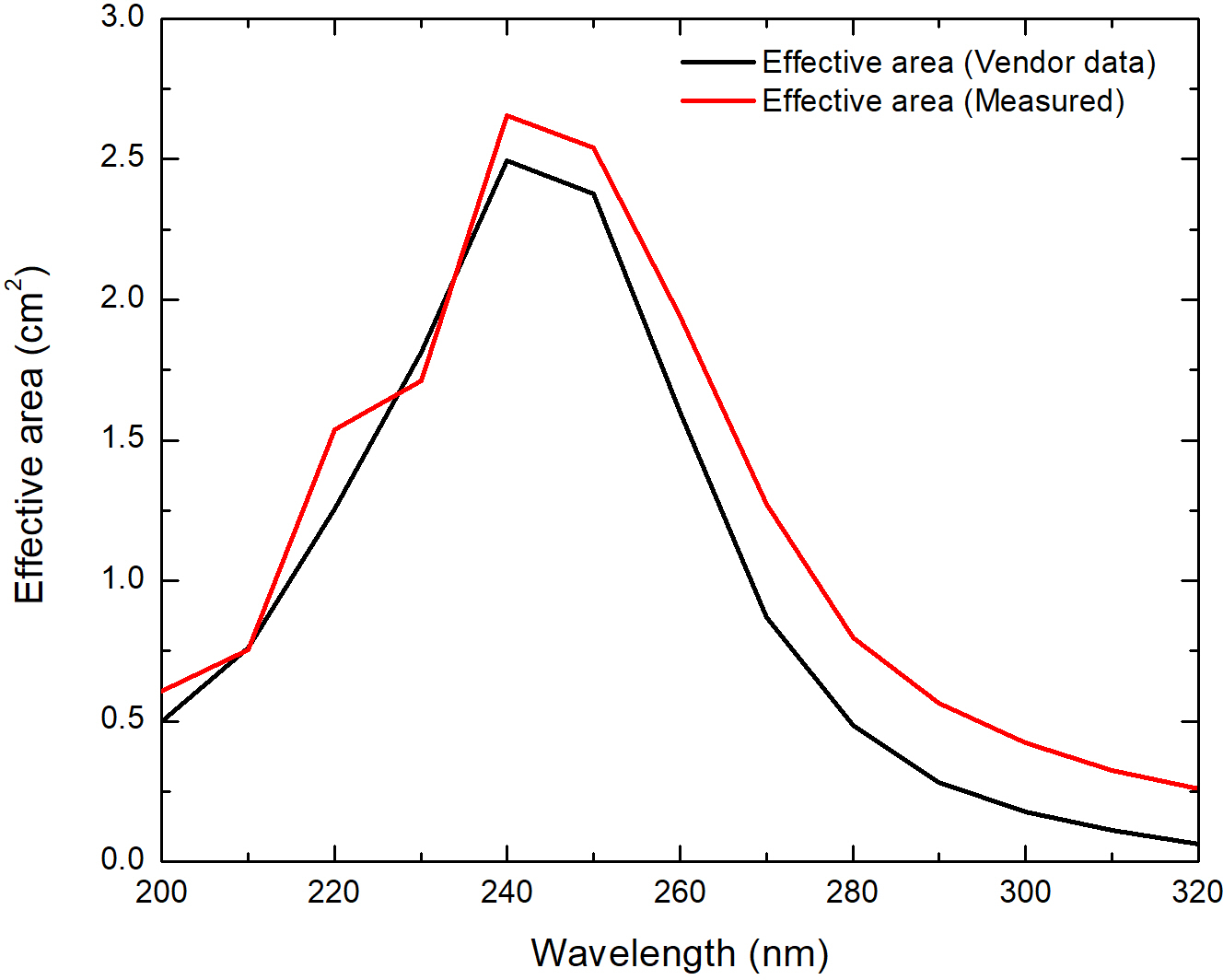} 
\caption{Effective area based on vendor data and as actually measured.}
\label{fig:effective area}
\end{figure}

We used a UV bandpass filter in LUCI to achieve the desired passband as well as to cut-off the long wavelengths, which we calibrated using the Acton spectrophotometer in 10$^{-4}$ mbar vacuum environment. The filter was mounted on the spectrophotometer's filter wheel which also has a blank filter, and we measured the transmission of both in 200--550 nm wavelength range with 10 nm step. The ratio of these measurements yielded the filter transmission, shown in Fig.~\ref{fig:Filter transmission}. 

\subsection{Effective area}

\begin{figure*}[h!]
\begin{center}
\includegraphics[scale=0.85]{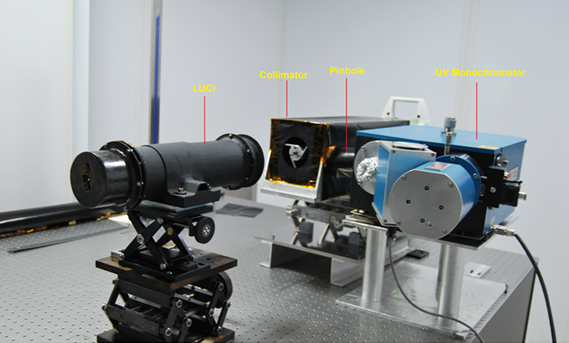}
\end{center}
\caption{ Optical bench set up to perform the PSF and effective area measurements.}
\label{fig:luci psf and effective area measurement set up}
\end{figure*}

We express the total system response, a constant characterizing the telescope’s efficiency in transmitting light, in terms of effective area, $A_{\rm eff}$ in cm$^2$. In paper~II, we have calculated it based on the manufacturers' data. 

Here, we measured the effective area of LUCI using the UV collimator/monochromator setup with a 20-mm circular mask and a NIST calibrated photodiode. The input UV flux was measured using a 20-mm diameter fused silica doublet lens to focus the light on the NIST calibrated photodiode. This lens we have previously calibrated using an Acton spectrophotometer. As with the QE measurement, we imaged the pinhole on the CCD for different wavelengths and measured the effective area (see Fig.~\ref{fig:luci psf and effective area measurement set up} for the experimental setup). The comparison of measured effective area and effective area based on vendor data is shown in Fig.~\ref{fig:effective area}. Earlier we have assumed the reflectivity of primary and secondary mirrors as 80\% (paper~II), however the final effective area measurements showed that the actual reflectivity in the 200--320 nm band is higher, yielding higher throughput.

We have updated the photometric calibration constants (calculated in paper~II), such as effective bandwidth, mean, pivot and effective wavelengths, and the total effective area. 

\begin{table*}[h!]
\begin{center}
\caption{Ground calibration values of photometric constants in nm.} 
\label{table:constants}
\begin{tabular}{|c|c|c|c|c|c|}
\hline
Effective Bandwidth $\D\l$ & Central $\l_0$ & Pivot $\l_{\rm p}$  &  \multicolumn{2}{|c|}{Effective $\lambda_{\rm eff}$}  \\
\hline
\multirow{2}{*}{57.15}   & \multirow{2}{*}{243.87}  & \multirow{2}{*}{242.4} &  Vega & HZ43  \\ 
\cline{4-5} 
  & & & 247.0 & 240.1  \\
\hline
\multicolumn{5}{l}{{\it Note}: \small Spectra of Vega and HZ43 white dwarf were obtained}\\
\multicolumn{5}{l}{\small from MAST IUE database (http://archive.stsci.edu/iue/).}
\end{tabular} 
\end{center}
\end{table*}

We estimate the effective bandwidth as the integral of the normalized effective area. This is equivalent to an ideal square filter with the same total area and average response as the actual filter. Effective bandwidth can be 10 times lower than raw bandwidth, or higher than the FWHM bandwidth and, as such, is better at comparing different bandpass filters than the raw bandpass. 

The mean (central) source-independent wavelength was calculated as 
\be
\l_{0}=\fr{\int \l A_{\rm norm}(\l) d\l}{\int A_{\rm norm}(\l) d\l}\,,
\label{eq:central}
\ee
where $A_{\rm norm}(\l)$ is the effective area (in cm$^2$) measured in the ground calibration normalized to 1. An exact relationship between $\la F_{\lambda}\ra$ and $\la F_{\nu}\ra$, the mean source intrinsic spectral-energy distribution in energy and frequency units, respectively, 
$\la F_{\l}\ra=\la F_{\nu}\ra c^2/\l^2_{\rm p}$, is provided by the pivot wavelength of the system,
\be
\l_{\rm p}=\sqrt{\fr{\int A_{\rm norm}(\l) \l d\l}{\int A_{\rm norm}(\l) d\l/\l}}\,,
\label{eq:pivot}
\ee
Both the central and the pivot wavelengths are independent of the spectrum of the source. The effect of the source power distribution over a given filter is included in the filter's effective wavelength
$\l_{\rm eff}$,
\be
\l_{\rm eff}= \fr{\int \l A(\l) F(\l) d\l}{\int A(\l) F(\l)d\l}\,,
\label{eq:effective}
\ee
where $F(\l)$ is the source spectrum in ergs/cm$^2$/s/\AA. This is the mean wavelength of the passband as weighted by the energy distribution of the source over the band and is especially useful in, for example, predicting the expected counts. 

\begin{figure}[h]
\begin{center}
\includegraphics[scale=0.32]{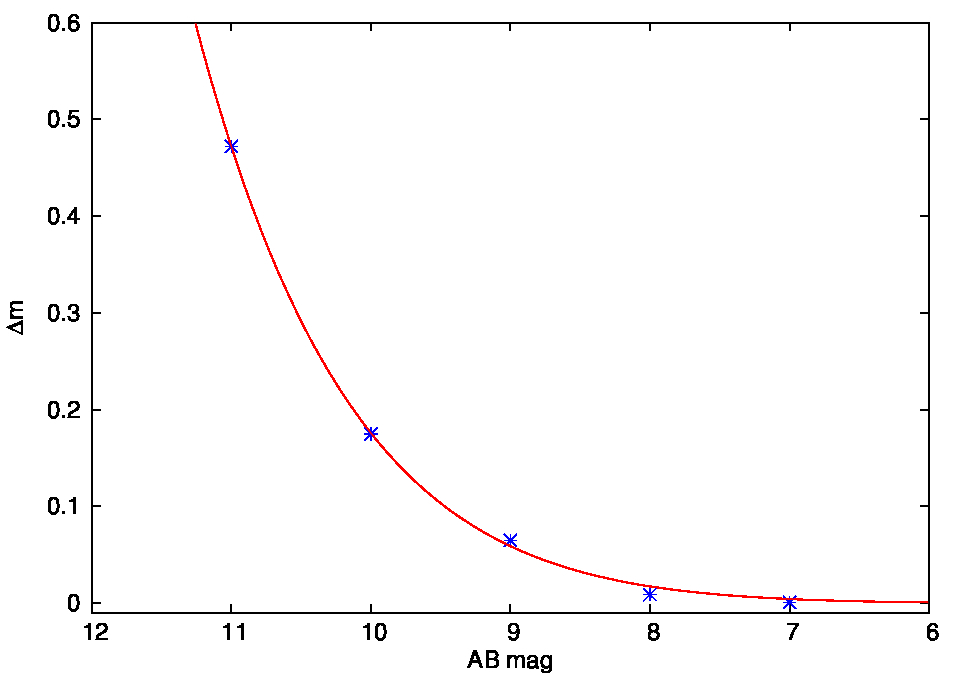} 
\end{center}
\caption{LUCI updated photometric accuracy. Stars are calculated data points and the curve is the best fit.}
\label{fig:LUCI Photometric accuracy}
\end{figure}

Keeping in mind the science objectives of detecting UV transients, we have also updated LUCI's sensitivity values following the method described in paper~II. We note that the sensitivity values (based on the calibration data) have not changed noticeably: with SNR of 4.3, LUCI can detect the source of $1.84\times 10^{-13}$ ergs/sec/cm$^2$/\AA\, flux density (which corresponds to the limiting magnitude of 12.7 AB). Therefore, LUCI sensitivity performance is satisfactory for the proposed science goals (see photometric accuracy in detecting the brightness variation Fig.~\ref{fig:LUCI Photometric accuracy}). The overall results of the ground calibrations are shown in Tables~\ref{table:constants} and \ref{table:CalibrationResults}. 

\begin{table}[h]
\caption{Ground calibrations results}
\begin{center}
\begin{tabular}{ll}
\hline 
{\bf Measured characteristics} &\\
\hline
\rule[-1ex]{0pt}{3.5ex} WFE & 53 nm \\
\rule[-1ex]{0pt}{3.5ex} Readout noise & 9 e  \\
\rule[-1ex]{0pt}{3.5ex} Dark current (@$23^{\circ}$C) & 1.2 e/px/sec \\
\rule[-1ex]{0pt}{3.5ex} Field of view & $27.55^{\prime}\times 20.37^{\prime}$ \\
\rule[-1ex]{0pt}{3.5ex} Plate scale & $1.21^{\prime\prime}$/pixel \\
\rule[-1ex]{0pt}{3.5ex} PSF (on-axis)& 3.85 pixels\\
\rule[-1ex]{0pt}{3.5ex} PSF (edge of field)& $4.25$ pixels\\
\rule[-1ex]{0pt}{3.5ex} Peak effective area (@240 nm) & 2.65 cm$^2$\\
\hline
{\bf Derived characteristics}&\\
\hline
\rule[-1ex]{0pt}{3.5ex} Total effective area & 153.8 cm$^2$\\
\rule[-1ex]{0pt}{3.5ex} Limiting magnitude & 12.7 AB\\
\rule[-1ex]{0pt}{3.5ex} Brightness limit & 2 AB\\
\hline
\end{tabular}
\label{table:CalibrationResults}
\end{center}
\end{table}

\section{Conclusions}

The LUCI payload has been assembled and calibrated in the class 1000 clean
room in the M.G.K. Menon Laboratory for Space Sciences, IIA. Performance
tests show that the assembled instrument meets the expected science
requirements. LUCI is stored in a container box with continuous purging
by high purity nitrogen to avoid any contamination to the optics. The
same container box will be used to transport the payload from the clean
room facility to the payload integration facility. The container and the
payload will be continuously purged with nitrogen until shortly before
the launch.

LUCI is to be mounted as a transit telescope on a lunar lander and will scan the NUV (200--320 nm) sky as the Moon rotates. The primary science goal is to observe bright UV transients (SNe, novae, TDE, etc.)---regime usually avoided by the traditional space UV telescopes due to the detectors safety concerns.

The journey of LUCI from our first contact with Team Indus until the final assembly has exposed us to both the opportunities and the pitfalls in these serendipitous flights. Because the opportunity arose through a chance meeting with a group of entrepreneurs, we did not have to go through a formal proposal round but, consequently, could not apply for separate funding through the normal channels, especially considering the level of uncertainty involved in a startup. Fortuitously, we had a long running program for instrument development which we used to build our instrument.

The goal of going to the Moon within the terms of the Google X-Prize was always ambitious, and the pressure of doing that within the given time frame and the cost cap made the actual achievement more difficult, both in terms of attracting the needed investment and in the realization of the hardware. The Google X-Prize was closed in March 2018, but all of the finalists are continuing with their missions, including the Team Indus.  

We are continuing with our space instruments development, and have several small payloads that are ready to fly. We have had serious discussions about flying them on our limited budget but have not secured a launch as yet. As launch costs decrease with the larger number of private players, we hope that we will have an opportunity to launch shortly.

\section{Acknowledgements}

We are grateful to the Team Indus for the flight opportunity and for the fruitful discussions regarding the LUCI payload. We thank all the staff at the M.G.K Menon laboratory (CREST) for helping us with storage and assembly of payload components in the clean room environment. Part of this research has been supported by the Department of Science and Technology (Government of India) under Grant EMR/2016/001450/PHY.

Some of the data presented in this paper were obtained from the Mikulski Archive for Space Telescopes (MAST). Space Telescope Science Institute (STScI) is operated by the Association of Universities for Research in Astronomy, Inc., under National Aeronautics and Space Administration (NASA) contract NAS5-26555. Support for MAST for non-Hubble Space Telescope (HST) data is provided by the NASA Office of Space Science via grant NNX09AF08G and by other grants and contracts.

\end{document}